\documentclass[aps,twocolumn,amssymb]{revtex4}
\usepackage{amsfonts,amssymb}
\usepackage{graphicx}
\usepackage{color}
\usepackage{textcomp}
\usepackage{ulem}
\begin{document}

\title{Mass spectra and Mott transitions of neutral mesons at finite temperature and magnetic field in frame of three-flavor Polyakov-extended Nambu-Jona-Lasino model}
\author{Luyang Li$^1$, Min Zhou$^2$, Zhiyang Liu$^2$, Chonglong Xie$^2$, Guoyun Shao$^2$}

\author{Shijun Mao$^2$}
 \email{maoshijun@mail.xjtu.edu.cn}
\affiliation{Xi'an University of Posts and Telecommunications, Xi'an, Shaanxi 710121, China$^1$\\
School of Physics, Xi'an Jiaotong University, Xi'an, Shaanxi 710049, China$^2$}

\begin{abstract}
Mass spectra and Mott transitions of neutral mesons $K_0, {\bar K}_0, \pi_0, \eta, \eta'$ at finite temperature and magnetic field are investigated in a three-flavor PNJL model. We focus on the effect of gluons, which is simulated by the Polyakov potential, and the inverse magnetic catalysis (IMC) effect, which is mimicked by using a magnetic field dependent parameter. Mass spectra show similar structure when introducing the contribution of Polyakov potential and IMC effect. The mass of $K_0\ ({\bar K}_0)$ meson $m_{K_0}=m_{{\bar K}_0}$ is controlled by chiral symmetry breaking and restoration. It increases with temperature in the low temperature region, and shows a mass jump at the Mott transition. Further increasing temperature, $m_{K_0}=m_{{\bar K}_0}$ firstly decreases and then increases with temperature. $\pi_0$ meson is not only the pseudo-Goldstone boson of chiral symmetry breaking, but also influenced by the flavor mixing of $\pi_0-\eta-\eta'$. The behavior of $m_{\pi_0}$ is different from $m_{K_0}=m_{{\bar K}_0}$ only at high temperature region, which decreases with temperature. For $\eta$ and $\eta'$ mesons, they are affected by both the $U_A(1)$ anomaly and the flavor mixing of $\pi_0-\eta-\eta'$. The mass of $\eta$ meson $m_{\eta}$ decreases with temperature in low temperature region and then shows a jump at its Mott transition. After that $m_{\eta}$ firstly decreases and later increases with temperature. $\eta'$ meson is a resonant state in vacuum and in medium, and its mass $m_{\eta'}$ continuously decreases and then increases with temperature. The mass jumps of $K_0, {\bar K}_0, \pi_0, \eta$ mesons are caused by the dimension reduction of the constituent quarks under external magnetic field. In the PNJL model, the Mott transition temperature of $K_0, {\bar K}_0, \pi_0$ mesons ($\eta$ meson) decreases (increases) with magnetic field. The IMC effect leads to no qualitative change to the meson Mott transition temperature but shifts them to the lower values.

\end{abstract}

\date{\today}

\maketitle

\section{Introduction}
	
In recent years, how the strong magnetic field affects the QCD phase transition and QCD matter has attracted much attention. Since it is widely believed that the strongest magnetic field in nature may be generated in the initial stage of relativistic heavy ion collisions. The initial magnitude of the magnetic field can reach $eB\sim (1-100)m_\pi^2$ in collisions at the Relativistic Heavy Ion Collider and the Large Hadron Collider~\cite{b0,b1,b2,b3,b4}, where $e$ is the electron charge and $m_\pi$ is the pion mass in vacuum.

On one side, QCD phase transitions under external magnetic field are widely investigated based on LQCD simulations and effective models. The LQCD simulations performed with physical pion mass report the inverse magnetic catalysis (IMC) phenomenon~\cite{lattice1,lattice2,lattice2sep,lattice4,lattice5,lattice6,lattice7,lattice9,rev1,rev2,rev3}. The quark chiral condensate (Polyakov loop) in high temperature region drops down (goes up) with increasing magnetic field, and the pseudo-critical temperature $T_{pc}$ of chiral symmetry restoration and deconfinement phase transitions decreases with increasing magnetic field. Many scenarios are proposed to understand this IMC phenomenon, but the physical mechanism is not clear~\cite{rev1,rev2,rev3,fukushima,mao,maosep1,maosep2,maosep3,maosep4,kamikado,bf1,lattice9,bf13,bf2,bf3,bf5,bf51,bf52,bf8,bf9,bf11,db1,db1sep,db2,db3,db5,db6,pnjl1,pnjl1sep,pnjl2,pnjl3,pnjl4,pqm,ferr1,ferr2,mhuang,meimao1,t0effect,meihuangmao,t0effectmao,maoxn2,maoxn3}.

On the other side, people study the magnetic field effects on the hadrons. As the Goldstone bosons of the chiral (isospin) symmetry breaking, the properties of neutral (charged) pions at finite magnetic field, temperature and density are widely investigated~\cite{c1,c3,hadron1,hadron2,qm1,qm2,sigma1,sigma2,sigma3,sigma4,l1,l2,l3,l4,lqcd5,lqcd6,ding2008.00493,njl2,meson,mfir,ritus5,ritus6,mao1,mao11,mao2,wang,coppola,phi,liuhao3,he,maocharge,maopion,yulang2010.05716,yulang2,q1,q2,q3,q4,huangamm1,huangamm2,q5,q6,q7,q8,q9,q10,feng,meijie3njl,ding2026}. Similar study of $K_0,{\bar K}_0$ ($K_\pm$) mesons are carried out under external magnetic field~\cite{ding2008.00493,su3meson1,q6,su3meson3,su3meson4,feng,meijie3njl,ding2026}, which is related to the chiral symmetry restoration phase transition (Kaon superfluid phase transition). Another interesting issue is $\eta,\ \eta'$ mesons, which is related to the $U_A(1)$ anomaly of QCD~\cite{ding2008.00493,su3meson1,q6,su3meson3,su3meson4,feng,meijie3njl}. Furthermore, there are some other works involving the $\rho$ meson and $\phi$ meson~\cite{Chernodub:2010qx,Chernodub:2011mc,Callebaut:2011uc,Ammon:2011je,Cai:2013pda,Frasca:2013kka,Andreichikov:2013zba,Liu:2014uwa,Liu:2015pna,Liu:2016vuw,Kawaguchi:2015gpt,Ghosh:2016evc,Ghosh:2017rjo,l1,Luschevskaya:2014mna,Luschevskaya:2015bea,lqcd5,Ding:2020jui,Ghosh,houphi}, heavy mesons~\cite{Marasinghe:2011bt,Machado:2013rta,Alford:2013jva,Machado:2013yaa,Cho:2014exa,Cho:2014loa,Dudal:2014jfa,Bonati:2015dka,Gubler:2015qok,Yoshida:2016xgm,Reddy:2017pqp,CS:2018mag} and baryons~\cite{Tiburzi:2008ma,Andreichikov:2013pga,Tiburzi:2014zva,Haber:2014zba,He:2016oqk,Deshmukh:2017ciw,Yakhshiev:2019gvb} in the magnetic field.

Our current work focuses on the mass spectra of neutral mesons $K_0, {\bar K}_0$, $\pi_0$, $\eta$ and $\eta'$ mesons at finite magnetic field and temperature, which are related to the chiral symmetry and $U_A(1)$ anomaly of QCD. When the magnetic field is of strength comparable with the strong interaction energy scale, such as $eB\sim m^2_\pi$, the quark structure of hadrons should be taken into account. We make use of the three flavor Polyakov-extended Nambu-Jona-Lasinio (PNJL) model~\cite{pnjl5,pnjl6,pnjl7,pnjl8,pnjl9,pnjl10,pnjl12}, which describes well the chiral restoration and deconfinement phase transitions and contains the $U_A(1)$ anomaly effect. In this model, quarks are treated in mean field level, gluon contribution is mainly simulated by the Polyakov potential, and mesons are the quantum fluctuations constructed from the quark bubble. Comparing with the previous studies~\cite{ding2008.00493,su3meson1,q6,su3meson3,su3meson4,feng,meijie3njl}, we consider the effect of gluons (simulated by the static background field) and the IMC effect (mimicked by the magnetic field dependent parameters) in this work. 
	
The rest of paper is arranged as follows. We introduce the magnetized $SU(3)$ PNJL model, and derive the formula for quark masses and neutral meson masses in Sec.\ref{form}. The numerical results and analysis of neutral meson masses at finite magnetic field and temperature are presented in Sec.\ref{numerical}. The summary is written in Sec.\ref{sum}.

\section{theoretical framework}
\label{form}


The three-flavor PNJL model under external magnetic field is defined with the Lagrangian density~\cite{pnjl5,pnjl6,pnjl7,pnjl8,pnjl9,pnjl10,pnjl12},
\begin{eqnarray}
\mathcal{L}&=&\bar{\psi}\left(i\gamma^{\mu}D_{\mu}-\hat{m}_0\right)\psi+\mathcal{L}_{ 4}+\mathcal{L}_{6}-{\cal U}(\Phi,\bar\Phi),\\
\mathcal{L}_{ 4}&=&G\sum_{\alpha=0}^{8}\left[(\bar{\psi}\lambda^{\alpha}\psi)^2+(\bar{\psi}i\gamma_5\lambda^{\alpha}\psi)^2\right],\nonumber \\
\mathcal{L}_{6}&=&-K\left[\text{det}\bar{\psi}(1+\gamma_5)\psi+\text{det}\bar{\psi}(1-\gamma_5)\psi \right],\nonumber\\
{\cal U}(\Phi,{\bar \Phi}) &=&  T^4 \left[-{b_2(t)\over 2} \bar\Phi\Phi -{b_3\over 6}\left({\bar\Phi}^3+\Phi^3\right)+{b_4\over 4}\left(\bar\Phi\Phi\right)^2\right].\nonumber
\label{lagrangian}
\end{eqnarray}
The covariant derivative $D^\mu=\partial^\mu+i Q A^\mu-i {\cal A}^\mu$ couples quarks to the two external fields, the magnetic field ${\bf B}=\nabla\times{\bf A}$ and the temporal gluon field  ${\cal A}^\mu=\delta^\mu_0 {\cal A}^0$ with ${\cal A}^0=g{\cal A}^0_a {\cal G}_a/2=-i{\cal A}_4$ in Euclidean space. The gauge coupling $g$ is combined with the $SU(3)$ gauge field ${\cal A}^0_a(x)$ to define ${\cal A}^\mu(x)$, and ${\cal G}_a$ are the Gell-Mann matrices in color space. We consider magnetic field ${\bf B}=(0, 0, B)$ along the $z$-axis by setting $A_\mu=(0,0,x B,0)$ in Landau gauge, which couples quarks of electric charge $Q=\text{diag}(Q_u,Q_d,Q_s)=\text{diag}(2/3 e,-1/3 e,-1/3 e)$. $\hat{m}_0=\text{diag}(m^u_0,m_0^d,m_0^s)$ is the current quark mass matrix in flavor space. The four-fermion interaction $\mathcal{L}_{4}$ represents the interaction in scalar and pseudo-scalar channels, with Gell-Mann matrices $\lambda^{\alpha},\ \alpha=1,2,...,8$ and $\lambda^0=\sqrt{2/3} \mathbf{I}$ in flavor space. The six-fermion interaction or Kobayashi-Maskawa-'t Hooft term $\mathcal{L}_{6}$ is related to the $U_A(1)$ anomaly~\cite{tHooft1,tHooft2,tHooft3,tHooft4,tHooft5}. The Polyakov potential ${\cal U}(\Phi,\bar\Phi)$ describes deconfinement, where $\Phi$ is the trace of the Polyakov loop $\Phi=\left({\text {Tr}}_c L \right)/N_c$, with $L({\bf x})={\cal P} \text {exp}[i \int^\beta_0 d \tau {\cal A}_4({\bf x},\tau)]= \text {exp}[i \beta {\cal A}_4 ]$ and $\beta=1/T$, the coefficient $b_2(t)=a_0+a_1 t+a_2 t^2+a_3 t^3$ with $t=T_0/T$ is temperature dependent, and the other coefficients $b_3$ and $b_4$ are constants. $T_0$ is the critical temperature of deconfinement phase transition in pure gauge LQCD.

\subsection{quarks}
When converting the six-fermion interaction into an effective four-fermion interaction in the mean field approximation, the Lagrangian density can be rewritten as
	\begin{eqnarray}
		\mathcal{L}&=&\bar{\psi}\left(i\gamma^{\mu}D_{\mu}-\hat{m}_0\right)\psi \\
		&+&\sum_{\alpha=0}^{8}\left[{\cal K}_\alpha^-\left(\bar{\psi}\lambda^\alpha\psi\right)^2+{\cal K}_\alpha^+\left(\bar{\psi}i\gamma_5\lambda^\alpha\psi\right)^2\right]\nonumber\\	&+&{\cal K}_{30}^-\left(\bar{\psi}\lambda^3\psi\right)\left(\bar{\psi}\lambda^0\psi\right)+{\cal K}_{30}^+\left(\bar{\psi}i\gamma_5\lambda^3\psi\right)\left(\bar{\psi}i\gamma_5\lambda^0\psi\right)\nonumber\\	&+&{\cal K}_{03}^-\left(\bar{\psi}\lambda^0\psi\right)\left(\bar{\psi}\lambda^3\psi\right)+{\cal K}_{03}^+\left(\bar{\psi}i\gamma_5\lambda^0\psi\right)\left(\bar{\psi}i\gamma_5\lambda^3\psi\right)\nonumber\\	&+&{\cal K}_{80}^-\left(\bar{\psi}\lambda^8\psi\right)\left(\bar{\psi}\lambda^0\psi\right)+{\cal K}_{80}^+\left(\bar{\psi}i\gamma_5\lambda^8\psi\right)\left(\bar{\psi}i\gamma_5\lambda^0\psi\right)\nonumber\\	&+&{\cal K}_{08}^-\left(\bar{\psi}\lambda^0\psi\right)\left(\bar{\psi}\lambda^8\psi\right)+{\cal K}_{08}^+\left(\bar{\psi}i\gamma_5\lambda^0\psi\right)\left(\bar{\psi}i\gamma_5\lambda^8\psi\right)\nonumber\\	&+&{\cal K}_{83}^-\left(\bar{\psi}\lambda^8\psi\right)\left(\bar{\psi}\lambda^3\psi\right)+{\cal K}_{83}^+\left(\bar{\psi}i\gamma_5\lambda^8\psi\right)\left(\bar{\psi}i\gamma_5\lambda^3\psi\right)\nonumber\\	&+&{\cal K}_{38}^-\left(\bar{\psi}\lambda^3\psi\right)\left(\bar{\psi}\lambda^8\psi\right)+{\cal K}_{38}^+\left(\bar{\psi}i\gamma_5\lambda^3\psi\right)\left(\bar{\psi}i\gamma_5\lambda^8\psi\right)\nonumber,
		\label{semilagrangian}
	\end{eqnarray}
	with the effective coupling constants
	\begin{eqnarray}
		\label{constants}
		&&{\cal K}_0^\pm=G\pm\frac{1}{3}K\left(\sigma_u+\sigma_d+\sigma_s\right),\\
		&&{\cal K}_1^\pm={\cal K}_2^\pm={\cal K}_3^\pm=G\mp\frac{1}{2}K\sigma_s,\nonumber\\
		&&{\cal K}_4^\pm={\cal K}_5^\pm=G\mp\frac{1}{2}K\sigma_d,\nonumber\\
		&&{\cal K}_6^\pm={\cal K}_7^\pm=G\mp\frac{1}{2}K\sigma_u,\nonumber\\
		&&{\cal K}_8^\pm=G\mp\frac{1}{6}K\left(2\sigma_u+2\sigma_d-\sigma_s\right),\nonumber\\
		&&{\cal K}_{03}^\pm={\cal K}_{30}^\pm=\mp\frac{1}{2\sqrt{6}}K\left(\sigma_u-\sigma_d\right),\nonumber\\
		&&{\cal K}_{08}^\pm={\cal K}_{80}^\pm=\mp\frac{\sqrt{2}}{12}K\left(\sigma_u+\sigma_d-2\sigma_s\right),\nonumber\\
		&&{\cal K}_{38}^\pm={\cal K}_{83}^\pm=\pm\frac{1}{2\sqrt{3}}K\left(\sigma_u-\sigma_d\right),	\nonumber
	\end{eqnarray}
	and chiral condensates
	\begin{eqnarray}
		\sigma_u=\langle\bar{u}u\rangle, \  \sigma_d=\langle\bar{d}d\rangle, \  \sigma_s=\langle\bar{s}s\rangle.
	\end{eqnarray}
	
%

The thermodynamic potential in mean field level contains the mean field part, Polyakov potential and quasi-particle part
	\begin{eqnarray}
		\label{omega1}
		\Omega &=&\sum_{f=u,d,s}\left(2 G\sigma_f^2\right)-4K\sigma_u\sigma_d\sigma_s+{\cal U}(\Phi,\bar\Phi)+\Omega_q, \\
		\Omega_q &=& - \sum_{f=u,d,s}\frac{|Q_f B|}{2\pi}\sum_{l}\alpha_l \int \frac{d p_z}{2\pi} \Bigg[3E_f \nonumber\\
		&+& T\ln\left(1+3\Phi e^{-\beta E_f^+}+3{\bar \Phi}e^{-2\beta E_f^+}+e^{-3\beta E_f^+}\right)\nonumber\\
&+& T\ln\left(1+3{\bar \Phi} e^{-\beta E_f^-}+3{ \Phi}e^{-2\beta E_f^-}+e^{-3\beta E_f^-}\right)\Bigg],\nonumber
	\end{eqnarray}
where $f=u,d,s$ means quark flavors, $l$ Landau levels, $\alpha_l=2-\delta_{l0}$ spin factor and $E_f^\pm=E_f \pm \mu_f$ contains quark energy $E_f=\sqrt{p^2_z+2 l |Q_f B|+m_f^2}$ of longitudinal momentum $p_z$ and effective quark masses $m_u=m_0^u-4G\sigma_u+2K\sigma_d\sigma_s$, $m_d=m_0^d-4G\sigma_d+2K\sigma_u\sigma_s$, $m_s=m_0^s-4G\sigma_s+2K\sigma_u\sigma_d$, and quark chemical potential $\mu_u=\frac{\mu_B}{3}+\frac{2\mu_Q}{3}$, $\mu_d=\frac{\mu_B}{3}-\frac{\mu_Q}{3}$, $\mu_s=\frac{\mu_B}{3}-\frac{\mu_Q}{3}+\frac{\mu_S}{3}$, with $\mu_B,\ \mu_Q\ ,\mu_S$ the chemical potential corresponding to the baryon number $B$, electric charge $Q$ and strangeness $S$, respectively.

The ground state at finite temperature, chemical potential and magnetic field is determined by minimizing the thermodynamic potential
\begin{eqnarray}
\label{gapeqs}
 \frac{\partial \Omega }{\partial \sigma_f}&=&0,\ (f=u,d,s),\nonumber\\
 \frac{\partial \Omega }{\partial \Phi}&=&0,\   \frac{\partial \Omega }{\partial {\bar \Phi}}=0.
\end{eqnarray}
The thermodynamic potential $\Omega$ is a function of order parameters (chiral condensates $\sigma_f$ and Polyakov loop $\Phi, {\bar \Phi}$), and hence we obtain five coupled gap equations. In the following calculations, we consider the case with vanishing quark chemical potential $\mu_{f=u,d,s}=0$, and therefore, we have $\Phi={\bar \Phi}$.

\subsection{mesons}
As quantum fluctuations above the mean field, mesons are constructed through quark bubble summation in the frame of random phase approximation (RPA)~\cite{njl2,njl1,njll2,njl3,njl4,njl5,zhuang}. Namely, the quark interaction via a meson exchange is effectively described by the Dyson-Schwinger equation,
\begin{equation}
{\cal M}_M(x,z)  = 2{\cal K} \delta(x-z)+\int d^4y\ 2{\cal K} \Pi_M(x,y) {\cal M}_M(y,z),
\label{dsequ}
\end{equation}
where ${\cal M}_M(x,y)$ represents the meson propagator from $x$ to $y$, ${\cal K}$ is the coupling constant between quarks, and the meson polarization function is the corresponding quark bubble,
\begin{equation}
\label{bubble}
\Pi_M(x,y) = i{\text {Tr}}\left[\Gamma_M^{\ast} S(x,y) \Gamma_M  S(y,x)\right]
\end{equation}
with the neutral meson vertex
\begin{equation}
		\Gamma_M=\left\{
		\begin{array}{l}
			i\gamma_5\lambda_0,\ \ M=\eta_0\\
			i\gamma_5\lambda_3,\ \ M=\pi_0\\
			i\gamma_5\lambda_8,\ \ M=\eta_8\ \ \\
i\gamma_5(\lambda_6\pm i\lambda_7)/\sqrt{2},\ \ M=K_0,\bar{K}_0
		\end{array}	\right. ,
	\end{equation}
the quark propagator matrix in flavor space $S=diag(S_u,\ S_d,\ S_s)$ at mean field level, and the trace taken in spin, color and flavor spaces.


For the neutral mesons, the meson momentum $k=(k_0,\vec{k})=(k_0, k_1,k_2,k_3)$ itself is conserved, and the meson propagator and the corresponding meson polarization function in momentum space are just the normal Fourier transformation of their expressions in coordinate space,
\begin{eqnarray}
\label{fourier1}
{\cal M}_{M}(k) &=&  \int d^4 (x-y) e^{i k \cdot (x-y)} {\cal M}_{M}(x,y),\\
\Pi_{M}(k) &=&  \int d^4 (x-y) e^{i k \cdot (x-y)} \Pi_{M}(x,y).
\end{eqnarray}
Considering the complete and orthogonal conditions of the plane wave $e^{-i k \cdot x}$, the neutral meson propagator in momentum space is simplified as
\begin{equation}
\label{npole}
{\cal M}_M(k)=\frac{2{\cal K}}{1-2{\cal K}\Pi_M(k)}.
\end{equation}
	
	The propagator of $K_0$ meson can be written as
	\begin{eqnarray}
		{\cal M}_{K_0}(k_0,\vec{k})=\frac{2{\cal K}_{6}^+}{1-2{\cal K}_{6}^+ \Pi_{K_0}(k_0,\vec{k})},
	\end{eqnarray}
Near the pole, we have
\begin{equation}
1-2 {\cal K}_{6}^+ \Pi_{K_0}(k)=\left(k_0^2-E^2_{K_0}({\vec k}^2)\right)\times {\text {const}},
\end{equation}
and the $K_0$ meson energy is given as
\begin{eqnarray}
E^2_{K_0}({\vec k}^2)=v^2_\parallel k^2_3+v^2_\perp (k^2_1+k^2_2) +m_{K_0}^2,		
	\end{eqnarray}
with different longitudinal and transverse velocity coefficients $v_{\parallel}, \ v_{ \perp}$ and pole mass $m_{K_0}$ of $K_0$ meson under external magnetic field.	

The pole mass $m_{K_0}$ is determined through the pole equation at zero momentum $\vec{k}=\vec{0}$,
	\begin{eqnarray}
		1-2{\cal K}_6^+ \Pi_{K_0}(m_{K_0},\vec{0})=0,
		\label{kaon}
	\end{eqnarray}
with the polarization function
	\begin{eqnarray}
		&\Pi_{K_0}(k_0,\vec{0})=\ \ \ \ \ \ \ \ \ \ \ \ \ \ \ \ \ \ \ \ \ \ \ \ \ \ \ \ \ \ \ \ \ \ \ \ \ \ \ \ \ \ \ \ \ \ \ \ \ \nonumber \\
		&\ \ \ \ J_1^{(d)}+J_1^{(s)}+\left[k_0^2-\left(m_s-m_d
		\right)^2\right] J_2^{(ds)}(k_0)\ \ \ \ \ \
		\label{pikaon}
	\end{eqnarray}
and
\begin{eqnarray}
		J_1^{(f)}&=&3\!\sum_{l}\alpha_l \frac{|Q_f B|}{2\pi}\!\int \frac{dp_z}{2\pi}\frac{1-2F_{\Phi}(E_f)}{E_f}, \\
		J_2^{(ds)}(k_0)&=&3\!\sum_{l}\!\alpha_l \frac{|Q_f B|}{2\pi}\int \frac{dp_z}{2\pi}\frac{1}{ E_s E_d}\ \ \ \ \\
		&\times&  \Bigg[\frac{E_s+E_d}{(E_s+E_d)^2-k^2_0}\left(1-F_{\Phi}(E_s)-F_{\Phi}(E_d)\right) \nonumber\\
		& +&\frac{E_s-E_d}{(E_s-E_d)^2-k^2_0} \left(F_{\Phi}(E_s)-F_{\Phi}(E_d) \right)  \nonumber \Bigg],\nonumber \\
        F_{\Phi}(x)&=&\frac{\Phi e^{-x/T}+2\Phi e^{-2x/T}+e^{-3x/T}}{1+3\Phi e^{-x/T}+3\Phi e^{-2x/T}+ e^{-3x/T}}.
		\label{kaonj2}	
	\end{eqnarray}
Because the $K_0$ meson is charge neutral, it is affected by the external magnetic field only through the constituent quarks. The formula for its propagator is the same as that without magnetic field except for the introduction of Landau levels in momentum integral. Moreover, when setting $\Phi=1$, the function $F_{\Phi}(x)$ becomes the Fermi-Dirac distribution, and we reproduce the analytical formula in the NJL model~\cite{meijie3njl}.

By interchanging two constituent quarks $d \leftrightarrow s$ in Eq.(\ref{pikaon}), the polarization function of $\bar{K}_0$ meson can be obtained. When the quark chemical potential is zero, $K_0$ and $\bar{K}_0$ mesons are degenerate and share the same mass. Therefore, in the following numerical results, we only show the results of $K_0$ meson.


The flavor mixing of $\pi_0-\eta-\eta'$ happens, since the magnetic field breaks the isospin symmetry between $u$ and $d$ quarks. Their meson propagators can be constructed in a matrix form with the RPA method,
	\begin{eqnarray}
		{\cal M}=2{\cal K}^+ (1-2 {\cal K}^+\Pi^P)^{-1},
	\end{eqnarray}
	where coupling constant ${\cal K}^+$ and polarization function $\Pi^P$ are $3\times3$ matrices
	\begin{eqnarray}
		{\cal K}^+={
			\left( \begin{array}{ccc}
				{\cal K}_0^+    & {\cal K}_{03}^+ & {\cal K}_{08}^+ \\
				{\cal K}_{30}^+ & {\cal K}_3^+    & {\cal K}_{38}^+ \\
				{\cal K}_{80}^+ & {\cal K}_{83}^+ & {\cal K}_8^+
			\end{array} \right)},\\
		\Pi^P={
			\left( \begin{array}{ccc}
				\Pi_0^P & \Pi_{03}^P & \Pi_{08}^P \\
				\Pi_{30}^P & \Pi_3^P & \Pi_{38}^P \\
				\Pi_{80}^P & \Pi_{83}^P & \Pi_8^P
			\end{array} \right)}.\
	\end{eqnarray}
	The matrix elements of coupling constant ${\cal K}^+$ are written in Eq.(\ref{constants}), and the elements of polarization function $\Pi^P$ are defined as
\begin{eqnarray}
		\Pi_{M'M}^P(k)=i \text{Tr}\left[\Gamma_{M'}^* S(p+\frac{1}{2}k)\Gamma_{M} S(p-\frac{1}{2}k)\right],
		\label{pimm}
	\end{eqnarray}
with index $3,0,8$ denoting $\pi_0, \eta_0, \eta_8$, respectively. For convenience of writing formula, we sometimes omit the argument $(k_0,\vec{k})$ in the polarization function and meson propagator.
	
	We can obtain the pole mass of $\pi_0,\ \eta,\ \eta'$ mesons by solving the equation at $\vec{k}=\vec{0}$
	\begin{eqnarray}
		\text{det}\left[{\cal M}^{-1}(k_0,\vec{0}) \right]=0.
\label{pieta}
	\end{eqnarray}
	The inverse of meson propagator matrix ${\cal M}$ is
	\begin{eqnarray}
		&&{\cal M}^{-1}=\frac{1}{2\text{det}{\cal K}^+}{
			\left( \begin{array}{ccc}
				\mathcal{A} & \mathcal{B} & \mathcal{C} \\
				\mathcal{B} & \mathcal{D} & \mathcal{E} \\
				\mathcal{C} & \mathcal{E} & \mathcal{F}
			\end{array} \right)},\ \\
		&&\mathcal{A}=\left({\cal K}_3^+ {\cal K}_8^+-{\cal K}_{38}^{+}{\cal K}_{38}^{+}\right) -2\Pi^P_0 \text{det}{\cal K}^+, \nonumber\\
		&&\mathcal{B}=\left({\cal K}_{38}^+ {\cal K}_{08}^+-{\cal K}_{8}^{+}{\cal K}_{03}^{+}\right) -2\Pi^P_{03} \text{det}{\cal K}^+, \nonumber\\
		&&\mathcal{C}=\left({\cal K}_{03}^+ {\cal K}_{38}^+-{\cal K}_{3}^{+}{\cal K}_{08}^{+}\right) -2\Pi^P_{08} \text{det}{\cal K}^+, \nonumber\\
		&&\mathcal{D}=\left({\cal K}_0^+ {\cal K}_8^+-{\cal K}_{08}^{+}{\cal K}_{08}^{+}\right) -2\Pi^P_3 \text{det}{\cal K}^+, \nonumber\\
		&&\mathcal{E}=\left({\cal K}_{03}^+ {\cal K}_{08}^+-{\cal K}_{0}^{+}{\cal K}_{38}^{+}\right) -2\Pi^P_{38} \text{det}{\cal K}^+, \nonumber\\
		&&\mathcal{F}=\left({\cal K}_3^+ {\cal K}_0^+-{\cal K}_{03}^{+}{\cal K}_{03}^{+}\right) -2\Pi^P_8 \text{det}{\cal K}^+ .\nonumber
	\end{eqnarray}
	with
	\begin{eqnarray}
		\label{polepieta}
		&&\Pi^P_0=\frac{2}{3}\left(\Pi^P_{uu}+\Pi^P_{dd}+\Pi^P_{ss}\right),\\
		&&\Pi^P_3=\Pi^P_{uu}+\Pi^P_{dd},\nonumber\\
		&&\Pi^P_8=\frac{1}{3}\left(\Pi^P_{uu}+\Pi^P_{dd}+4\Pi^P_{ss}\right),\nonumber\\
		&&\Pi^P_{03}=\Pi^P_{30}=\frac{\sqrt{6}}{3}\left(\Pi^P_{uu}-\Pi^P_{dd}\right),\nonumber\\
		&&\Pi^P_{08}=\Pi^P_{80}=\frac{\sqrt{2}}{3}\left(\Pi^P_{uu}+\Pi^P_{dd}-2\Pi^P_{ss}\right),\nonumber\\
		&&\Pi^P_{38}=\Pi^P_{83}=\frac{\sqrt{3}}{3}\left(\Pi^P_{uu}-\Pi^P_{dd}\right),\nonumber
	\end{eqnarray}
	and
	\begin{eqnarray}
		\Pi^P_{ff}(k_0)&=&J_1^{(f)}+k^2_0 J_2^{(ff)}(k_0),\\
		J_2^{(ff)}(k_0)&=&3\!\sum_{l}\!\alpha_l \frac{|Q_f B|}{2\pi}\!\!\int \frac{dp_z}{2\pi}\frac{1-2F_{\Phi}(E_f)}{E_f\!\left(\!4E_f^2-k_0^2\right)}.\nonumber
	\end{eqnarray}

The formula for each element in polarization function matrix $\Pi^P_{ff}$ is the same as that without magnetic field except for the introduction of Landau levels in momentum integral. When setting $\Phi=1$, the function $F_{\Phi}(x)$ becomes the Fermi-Dirac distribution, and we reproduce the analytical results in the NJL model~\cite{meijie3njl}.

The flavor mixing of $\pi_0-\eta-\eta'$ mesons caused by the magnetic field is represented by the non-vanishing elements ${\cal K}_{03}^\pm$, ${\cal K}_{30}^\pm$, ${\cal K}_{38}^\pm$ and ${\cal K}_{83}^\pm$ in coupling constants, and the non-vanishing elements $\Pi^P_{03},\ \Pi^P_{30},\ \Pi^P_{38},\ \Pi^P_{83}$ in polarization function matrix. In case of neglecting $U_A(1)$ anomaly effect by setting vanishing six-fermion interaction with $K=0$, we have ${\cal K}_{03}^\pm={\cal K}_{30}^\pm={\cal K}_{38}^\pm={\cal K}_{83}^\pm=0$. The flavor mixing of $\pi_0-\eta-\eta'$ mesons still happens, because the polarization function matrix has non-vanishing elements $\Pi^P_{03},\ \Pi^P_{30},\ \Pi^P_{38},\ \Pi^P_{83}$ under finite magnetic field.

\subsection{parameters}
Because of the contact interaction between quarks, the ultraviolet divergence cannot be eliminated through renormalization, and a proper regularization scheme is needed. In our work, we apply the Pauli-Villars regularization~\cite{njl2,mao1,mao11,mao2,maopion,meijie3njl}, which is gauge invariant and can guarantee the law of causality at finite magnetic field. By fitting the physical quantities, pion mass $m_{\pi}=138\ \text{MeV}$, pion decay constant $f_{\pi}=93\ \text{MeV}$, kaon mass $m_K=495.7\ \text{MeV}$, $\eta'$ meson mass $m_{\eta\prime}=957.5\ \text{MeV}$ in vacuum, we fix the current masses of light quarks $m_0^{u}=m_0^{d}=5.5\ \text{MeV}$, and obtain the parameters $m_0^s=154.7\ \text{MeV}$, $G\Lambda^2=3.627$, $K\Lambda^5=92.835$, $\Lambda=1101\ \text{MeV}$. For the Polyakov potential, the parameters are chosen as~\cite{pnjl6} $a_0=6.75$, $a_1=-1.95$, $a_2=2.625$, $a_3=-7.44$, $b_3=0.75$, $b_4=7.5$, and $T_0=270$ MeV.

\begin{figure}[htb]
\begin{center}
\includegraphics[width=7cm]{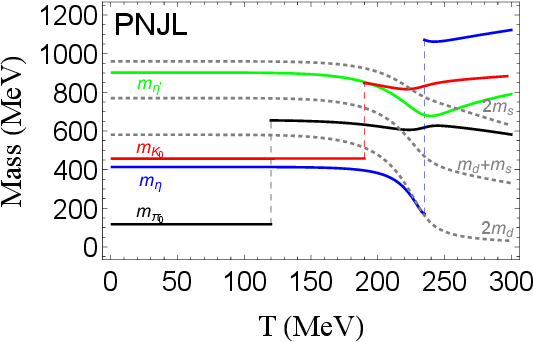}
\includegraphics[width=7cm]{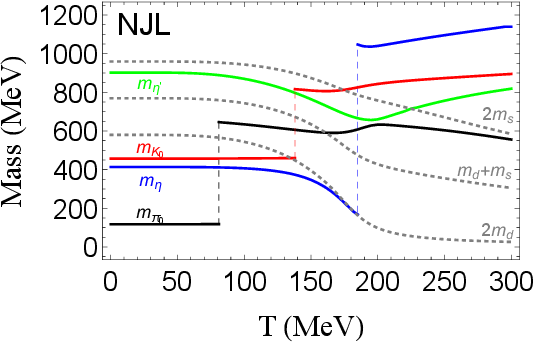}
\end{center}
\caption{ Mass spectra of $K_0$ meson $m_{K_0}$ (red solid lines), $\pi_0$ meson $m_{\pi_0}$ (black solid lines), $\eta$ meson $m_{\eta}$ (blue solid lines), $\eta'$ meson $m_{\eta'}$ (green solid lines) and quark mass sum $m_d+m_s$, $2m_d$, $2m_s$ (gray dotted lines) at finite temperature with fixed magnetic field $eB=20m_{\pi}^2$ and vanishing quark chemical potential $\mu_f=0$ in the PNJL model (upper panel) and in the NJL model (lower panel).}
\label{figmass}
\end{figure}
\begin{figure}[htb]
\begin{center}
\includegraphics[width=7cm]{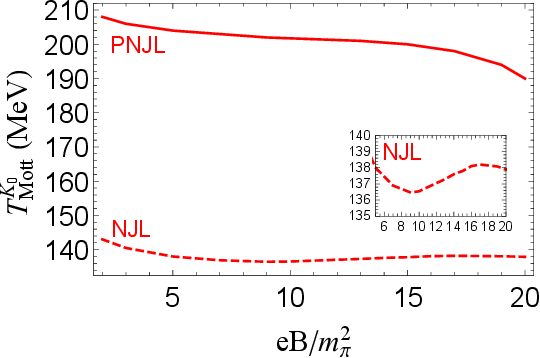}
\includegraphics[width=7cm]{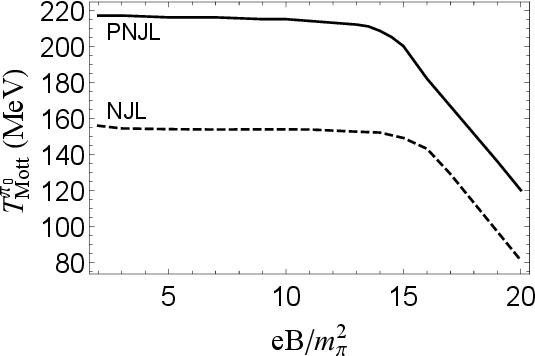}
\includegraphics[width=7cm]{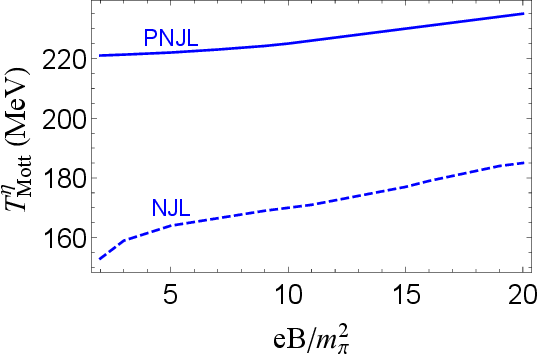}
\end{center}
\caption{Mott transition temperature of $K_0$ meson (upper panel), $\pi_0$ meson (middle panel) and $\eta$ meson (lower panel) as a function of magnetic field in PNJL model (solid lines) and in NJL model (dashed lines).}
\label{figmott}
\end{figure}

\section{Numerical Results}
\label{numerical}
\subsection{gluon effect}
Fig.\ref{figmass} plots the mass spectra of $K_0$ meson $m_{K_0}$ (red solid lines), $\pi_0$ meson $m_{\pi_0}$ (black solid lines), $\eta$ meson $m_{\eta}$ (blue solid lines), $\eta'$ meson $m_{\eta'}$ (green solid lines) at finite temperature with fixed magnetic field $eB=20m_{\pi}^2$ and vanishing quark chemical potential $\mu_f=0$ in the PNJL model (upper panel) and in the NJL model (lower panel). To demonstrate the chiral symmetry restoration and meson Mott transition, we also show the results of quark mass sum $m_d+m_s$, $2m_d$ and $2m_s$ (gray dotted lines).
	
As pseudo-Goldstone boson of chiral symmetry spontaneous breaking, when the constituent quark mass decreases, the mass of $K_0$ meson will generally increase. Therefore chiral symmetry restoration is expected to induce an intersection between the $K_0$ meson mass $m_{K_0}$ and the sum of two constituent quark masses $m_d+m_s$, which defines the Mott transition of $K_0$ meson~\cite{mott1,zhuang,mott2,mott3}. Based on the polarization function (\ref{pikaon}), with $p_z=0$ and lowest Landau level $l=0$, the integral term $\frac{1}{(E_s+E_d)^2-k^2_0}$ diverges when $k_0=m_d+m_s$. This infrared divergence leads to the mass jump of $K_0$ meson at the Mott transition. As shown in Fig.\ref{figmass} upper panel, with fixed magnetic field $eB=20m^2_{\pi}$, the mass of $K_0$ meson $m_{K_0}$ (see red lines) increases with temperature in the low temperature region. And then, it shows an abrupt jump at the Mott transition temperature $T^{K_0}_{\text {Mott}}=190$ MeV, where the $K_0$ meson changes from the bound state with $m_{K_0}<m_d+m_s$ to the resonant state $m_{K_0}>m_d+m_s$. After that, $m_{K_0}$ firstly decreases and then increases with temperature.

It should be mentioned that, when the $K_0$ meson mass $m_{K_0}$ reaches the sum of the effective masses of constituent quarks corresponding to higher Landau levels $\sqrt{2 l |Q_d B|+m_d^2}+\sqrt{2 l |Q_s B|+m_s^2}$ with $l=1,2,...$, such infrared divergence still exists. Therefore, when the mass of $K_0$ meson $m_{K_0}$ is large enough, it may undergo more than one jump, which is, however, not observed in the current numerical calculations. Besides, the integral term $\frac{1}{(E_s-E_d)^2-k^2_0}$ also has infrared divergence with small value of $k_0$, which is not relevant for the $K_0$ meson mass.

Under external magnetic field, $\pi_0$ meson is not only the pseudo-Goldstone boson of chiral symmetry breaking, but also influenced by the flavor mixing of $\pi_0-\eta-\eta'$. As shown in the black lines of Fig.\ref{figmass} upper panel, the  mass of $\pi_0$ meson $m_{\pi_0}$ increases with temperature, and a mass jump from $m_{\pi_0}<2m_d$ to $2m_d<m_{\pi_0}<2m_s$ happens at its Mott transition temperature $T^{\pi_0}_{\text {Mott}}=120$ MeV. Further increasing temperature, $m_{\pi_0}$ slightly decreases, then increases and finally decreases with temperature, showing a non-monotonic behavior. The different behavior between $\pi_0$ meson and $K_0$ meson at high temperature region is due to the flavor mixing effect.

For $\eta$ and $\eta'$ mesons, they are affected by both the $U_A(1)$ anomaly and the flavor mixing of $\pi_0-\eta-\eta'$. The mass of $\eta$ meson $m_{\eta}$ (see the blue lines of Fig.\ref{figmass} upper panel) decreases with temperature in low temperature region and then jumps from $m_{\eta}<2m_d$ to $m_{\eta}>2m_s$ at its Mott transition temperature $T^{\eta}_{\text {Mott}}=235$ MeV. After that $m_{\eta}$ firstly decreases and later increases with temperature. $\eta'$ meson is a resonant state in vacuum and in medium (see the green lines of Fig.\ref{figmass} upper panel), which has the mass larger than two times of the $d$-quark. With the increase of temperature, the mass of $\eta'$ meson $m_{\eta'}$ continuously decreases and then increases with temperature.

Each matrix element in polarization function matrix Eq.(\ref{polepieta}) shows infrared divergence at $p_z=0$ and $k_0=2\sqrt{m_f^2+2l|Q_f B|}$ with $f=u,d,s,\ l=0,1,2,...$, due to the integral term $\frac{1}{4E_f^2-k_0^2}$. Although the mixing of $\pi_0-\eta-\eta'$ meson exists, the relevant infrared divergence for the Mott transition of $\pi_0$ ($\eta$) meson lies in the lowest Landau level, where we observe the mass jump of $\pi_0$ ($\eta$) meson from $m_{\pi_0}<2m_d$ ($m_{\eta}<2m_d$) to $m_{\pi_0}>2m_u$ ($m_{\eta}>2m_s$). $\eta'$ meson is always in a resonant state without mass jump and Mott transition.

Fig.\ref{figmott} solid lines plot the Mott transition temperature of $K_0,\ \pi_0,\ \eta$ mesons as a function of magnetic field in PNJL model. The Mott transition temperature of $K_0$ meson $T^{K_0}_{\text {Mott}}$ decreases with magnetic field, showing about $10\%$ reduction when $eB=20m^2_{\pi}$. The Mott transition temperature of $\pi_0$ meson $T^{\pi_0}_{\text {Mott}}$ slightly decreases in weak magnetic field region, and goes down fast in strong magnetic field region. The Mott transition temperature of $\eta$ meson $T^{\eta}_{\text {Mott}}$ increases with magnetic field, with less than $10\%$ enhancement up to $eB=20m^2_{\pi}$.

In the end of this part, we make comparison between the results in the PNJL model and the NJL model. The mass spectra of $K_0,\ \pi_0,\ \eta, \ \eta'$ mesons from PNJL model in Fig.\ref{figmass} upper panel and from NJL model in Fig.\ref{figmass} lower panel show similar structure. At $T=0$, it is easy to analytically prove that the meson masses are the same in PNJL and NJL models due to the decoupling of the Polyakov loop with the quarks. At low and high temperature region, their values are slightly different. In the medium temperature region, the Mott transitions occurs. The NJL model (Fig.\ref{figmott} dashed lines) and PNJL model (Fig.\ref{figmott} solid lines) give qualitatively similar results for Mott transition temperature of $\pi_0,\ \eta$ mesons. However, for $K_0$ meson, $T^{K_0}_{\text {Mott}}$ in NJL model shows non-monotonic behavior in strong magnetic field region, which is different from the monotonic decreasing behavior in PNJL model. The value of Mott transition temperatures in PNJL are larger than in NJL model. This quantitative different can be reduced by using smaller value for parameter $T_0$ in the Polyakov potential ${\cal U}(\Phi,\bar\Phi)$.

\begin{figure}[htb]
\begin{center}
\includegraphics[width=6.9cm]{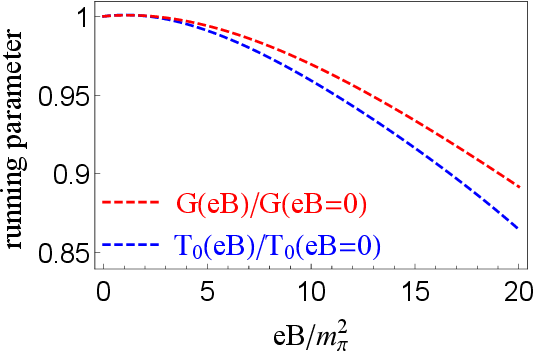}\\
\includegraphics[width=7cm]{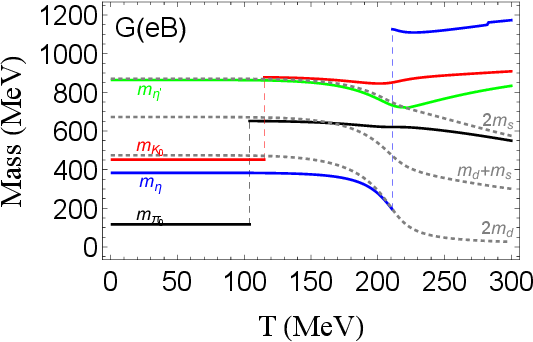}\\
\includegraphics[width=7cm]{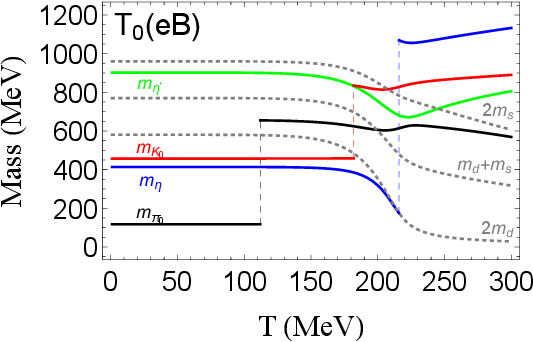}
\end{center}
\caption{(upper panel) Magnetic field dependent parameters $G(eB)$ (red line) and $T_0(eB)$ (blue line) fitted from LQCD reported decreasing pseudocritical temperature of chiral restoration phase transition $T_{pc}^c(eB)/T_{pc}^c(eB=0)$ under external magnetic field~\cite{lattice1}. (middle and lower panel) Mass spectra of $K_0$ meson $m_{K_0}$ (red solid lines), $\pi_0$ meson $m_{\pi_0}$ (black solid lines), $\eta$ meson $m_{\eta}$ (blue solid lines), $\eta'$ meson $m_{\eta'}$ (green solid lines) and quark mass sum $m_d+m_s$, $2m_d$, $2m_s$ (gray dotted lines) at finite temperature with fixed magnetic field $eB=20m_{\pi}^2$ and vanishing quark chemical potential $\mu_f=0$ in the PNJL model with IMC effect in $G(eB)$ scheme (middle panel) and with IMC effect in $T_0(eB)$ scheme (lower panel).}
\label{figmassimc}
\end{figure}

\subsection{IMC effect}
By fitting the LQCD reported decreasing pseudocritical temperature of chiral symmetry restoration $T_{pc}^c(eB)/T_{pc}^c(eB=0)$ under external magnetic field~\cite{lattice1}, we introduce the IMC effect in our three-flavor PNJL model through a magnetic field dependent parameter $G(eB)$~\cite{bf8,bf9,maoxn2,Liu:2016vuw,yulang2,su3meson4,maoxn3} and $T_0(eB)$~\cite{t0effectmao,t0effect,pnjl3,maoxn2,maoxn3}, respectively, which represent the influence of external magnetic field to the quark-gluon interaction. On one side, the coupling between quarks plays a significant role in determining the spontaneous breaking and restoration of chiral symmetry. Considering the direct interaction between quarks and external magnetic field, a magnetic field dependent coupling $G(eB)$~\cite{bf8,bf9,maoxn2,Liu:2016vuw,yulang2,su3meson4,maoxn3} is introduced into the PNJL model. On the other side, the interaction between the Polyakov loop and sea quarks may be important for the mechanism of IMC~\cite{lattice9}. A magnetic field dependent Polyakov loop scale parameter $T_0(eB)$~\cite{t0effectmao,t0effect,pnjl3,maoxn2,maoxn3} is introduced into the PNJL model to mimic the reaction of the gluon sector to the presence of magnetic fields. As plotted in Fig.\ref{figmassimc} upper panel, $G(eB)$ and $T_0(eB)$ are both monotonic decreasing functions of magnetic field~\cite{maoxn3}. We have checked that, with our fitted parameter $G(eB)$ or $T_0(eB)$, the increase (decrease) of chiral condensates with magnetic fields at the low (high) temperature, the increase of Polyakov loop with magnetic fields in the whole temperature region and the reduction of pseudocritical temperature of deconfinement phase transition under magnetic fields can be realized.

We calculate the mass spectra of $K_0,\ \pi_0,\ \eta, \ \eta'$ mesons and their Mott transition temperature based on the new PNJL model including IMC effect. As shown in Fig.\ref{figmassimc} middle (lower) panel, the mass spectra of $K_0,\ \pi_0,\ \eta, \ \eta'$ mesons are plotted as functions of temperature with fixed magnetic field $eB=20m^2_{\pi}$ in $G(eB)$ ($T_0(eB)$) scheme to consider the IMC effect. Qualitatively, they show similar structure as in the case neglecting IMC effect (see Fig.\ref{figmass} upper panel). Quantitatively, IMC effect slightly modifies the meson masses and leads to a lower value for meson Mott transition temperature, see Table.\ref{table3pnjl}.\\


\begin{table}
\begin{tabular}{cccc}
\hline
\hline
                      & $T^{\pi_0}_{\text {Mott}}$ \ & $T^{K_0}_{\text {Mott}}$\  & $T^{\eta}_{\text {Mott}}$  \\
                      & ({\text {MeV}})    & ({\text {MeV}})  & ({\text {MeV}})\\
\hline
$\text{w/o \ IMC} $   & 120                & 190              & 235 \\
\hline
$G(eB)$               & 104                & 115              & 211 \\
\hline
$T_0(eB)$             & 112                & 182              & 216 \\
\hline
\hline
\end{tabular}
\caption{Results of Mott transition temperature of $\pi_0$, $K_0$ and $\eta$ mesons with fixed magnetic field ($eB=20m^2_{\pi}$) in PNJL model without IMC effect and with IMC effect in $G(eB)$ and $T_0(eB)$ scheme.}
\label{table3pnjl}
\end{table}

\section{summary}
\label{sum}
Mass spectra and Mott transitions of neutral mesons $K_0, {\bar K}_0, \pi_0, \eta, \eta'$ at finite temperature and magnetic field are investigated in a three-flavor PNJL model. On one hand, we consider the effect of gluons, which is simulated by the Polyakov potential, by making comparison between the results of PNJL model and NJL model. On the other hand, we study the inverse magnetic catalysis (IMC) effect through the magnetic field dependent parameter in the PNJL model, which modifies the quark interaction ($G(eB)$ scheme) and Polyakov potential ($T_0(eB)$ scheme), respectively.

Mass spectra show similar structure when introducing the contribution of Polyakov potential and IMC effect. The mass of $K_0,{\bar K}_0$ meson $m_{K_0}=m_{{\bar K}_0}$ is controlled by chiral symmetry breaking and restoration. It increases with temperature in the low temperature region, and shows a mass jump at the Mott transition. Further increasing temperature, $m_{K_0}=m_{{\bar K}_0}$ firstly decreases and then increases with temperature. $\pi_0$ meson is not only the pseudo-Goldstone boson of chiral symmetry breaking, but also influenced by the flavor mixing of $\pi_0-\eta-\eta'$. The behavior of $m_{\pi_0}$ is different from $m_{K_0}$ only at high temperature region, which decreases with temperature. For $\eta$ and $\eta'$ mesons, they are affected by both the $U_A(1)$ anomaly and the flavor mixing of $\pi_0-\eta-\eta'$. The mass of $\eta$ meson $m_{\eta}$ decreases with temperature in low temperature region and then shows a jump at its Mott transition. After that $m_{\eta}$ firstly decreases and later increases with temperature. $\eta'$ meson is a resonant state in vacuum and in medium, and its mass $m_{\eta'}$ continuously decreases and then increases with temperature. Including the gluon effect and IMC effect only modifies the value of meson masses.

The mass jumps of $K_0, {\bar K}_0, \pi_0, \eta$ mesons at their Mott transitions are caused by the dimension reduction of the constituent quarks under external magnetic field. Introducing the gluon effect leads to some changes on their Mott transition temperatures. In the PNJL model, the Mott transition temperature of $K_0, {\bar K}_0, \pi_0$  mesons ($ \eta$ meson) decreases (increases) with magnetic field. However, in the NJL model, the Mott transition temperature of $K_0 ({\bar K}_0)$ meson shows non-monotonic behavior in strong magnetic field region. The IMC effect, simulated both in $G(eB)$ and $T_0(eB)$ scheme, leads to no qualitative difference but the lower value for meson Mott transition temperature.

It should be mentioned that $K_0,\ \pi_0,\ \eta$ mesons are in the resonant state with finite width after the mass jump, and $\eta'$ meson is in the resonant state in the whole temperature region. In principle, we should solve the pole equations (\ref{kaon}) and (\ref{pieta}) with the complex meson mass $m_M+i{ \gamma}_M$. Considering much smaller width ${\gamma}_M$ than mass $m_M$ for mesons, we use the pole equations (\ref{kaon}) and (\ref{pieta}) to derive meson masses in the whole temperature regions for simplicity~\cite{zhuang,meijie3njl}, which will not change our conclusions for meson mass spectra and Mott transitions.\\


\noindent {\bf Acknowledgement:} Shijun Mao is supported by the National Natural Science Foundation of China under Grant No.12275204. Guoyun Shao is supported by the National Natural Science Foundation of China under Grant No.12475145 and Natural Science Basic Research Plan in Shaanxi Province of China (Program No. 2024JC-YBMS-018). \\



\begin{thebibliography}{99}

\bibitem{b0} J. Rafelski and B. Muller, Phys. Rev. \textbf{Lett}. \textbf{36}, 517 (1976).
\bibitem{b1} D. E. Kharzeev, L. D. McLerran and H. J. Warringa, Nucl. Phys. \textbf{A}\textbf{803}, 227 (2008).
\bibitem{b2} V. Skokov, A. Y. Illarionov and T. Toneev, Int. J. Mod. Phys. \textbf{A}\textbf{24}, 5925 (2009).
\bibitem{b3} W. T. Deng and X. G. Huang, Phys. Rev. \textbf{C}\textbf{85}, (2012) 044907; Phys. Lett. \textbf{B}\textbf{742}, 296 (2015).
\bibitem{b4} K. Tuchin, Adv. High Energy Phys. 490495 (2013).

\bibitem{lattice1} G. S. Bali, F. Bruckmann, G. Endr$\ddot{o}$di, Z. Fodor, S. D. Katz, S. Krieg, A. Schaefer and K. K. Szabo, J. High Energy Phys. {\bf 02}, (2012) 044.
\bibitem{lattice2} G. S. Bali, F. Bruckmann, G. Endr$\ddot{o}$di, Z. Fodor, S. D. Katz and A. Schaefer, Phys. Rev. {\bf D86}, 071502 (2012).
\bibitem{lattice2sep} G. S. Bali, F. Bruckmann, G. Endr$\ddot{o}$di, Z. Fodor, S. D. Katz and A. Schaefer, J. High Energy Phys. {\bf 08}, 177 (2014).

\bibitem{lattice9} F. Bruckmann, G. Endr$\ddot{o}$di and T. G. Kovacs, J. High Energy Phys. {\bf 04}, 112 (2013).
\bibitem{lattice4} V. G. Bornyakov, P. V. Buividovich, N. Cundy, O. A. Kochetkov and A. Schaefer, Phys. Rev. {\bf D90}, 034501 (2014).
\bibitem{lattice5} G. Endr$\ddot{o}$di, J. High Energy Phys. {\bf 07}, 173 (2015).
\bibitem{lattice6} G. Endr$\ddot{o}$di, M. Giordano, S. D. Katz, T. G. Kovacs and F. Pittler, J. High Energy Phys. {\bf 07}, 009 (2019).
\bibitem{lattice7} H. T. Ding, S. T. Li, J. H. Liu and X. D. Wang, Phys. Rev. {\bf D105}, 034514 (2022).


\bibitem{rev1}K. Hattori, K. Itakura and S. Ozaki, Prog. Part. Nucl. Phys. {\bf 133}, 104068 (2023).
\bibitem{rev2}G. Endr$\ddot{o}$di, Prog. Part. Nucl. Phys. {\bf 141}, 104153 (2025).
\bibitem{rev3}A. Yamamoto, Eur. Phys. J. {\bf A57}, 211 (2021).

\bibitem{fukushima} K. Fukushima and Y. Hidaka, Phys. Rev. Lett {\bf 110}, 031601 (2013).
\bibitem{mao} S. J. Mao, Phys. Lett. {\bf B758}, 195 (2016).
\bibitem{maosep1}  S. J. Mao, Phys. Rev. {\bf D94}, 036007 (2016).
\bibitem{maosep2}  S. J. Mao, Phys. Rev. {\bf D97}, 011501(R) (2018).
\bibitem{maosep3}  S. J. Mao, Chin. Phys. {\bf C45}, 021004 (2021).
\bibitem{maosep4}  S. J. Mao, Phys. Rev. {\bf D106}, 034018 (2022).
\bibitem{kamikado}  K. Kamikado and T. Kanazawa, J. High Energy Phys. {\bf 03}, 009 (2014).
\bibitem{bf1} J. Y. Chao, P. C. Chu and M. Huang, Phys. Rev. {\bf D88}, 054009 (2013).

\bibitem{bf13} J. Braun, W. A. Mian and S. Rechenberger, Phys. Lett. {\bf B755}, 265 (2016).
\bibitem{bf2} N. Mueller and J. M. Pawlowski,  Phys. Rev. {\bf D91}, 116010 (2015).
\bibitem{bf3} T. Kojo and N. Su, Phys. Lett. {\bf B720}, 192 (2013).
\bibitem{bf5} A. Ayala, M. Loewe, A. J. Mizher and R. Zamora, Phys. Rev. {\bf D90}, 036001 (2014).
\bibitem{bf51}A. Ayala, L. A. Hernandez, A. J. Mizher, J. C. Rojas and C. Villavicencio, Phys. Rev. {\bf D89}, 116017 (2014).
\bibitem{bf52}A. Ayala, C. A. Dominguez, L. A. Hernandez, M. Loewe and R. Zamora, Phys. Rev. {\bf D92}, 096011 (2015).
\bibitem{bf8} R. L. S. Farias, K. P. Gomes, G. Krein, and M. B. Pinto, Phys. Rev. {\bf C90}, 025203 (2014).
\bibitem{bf9} M. Ferreira, P. Costa, O. Lourenco, T. Frederico, and C. Provid$\hat e$ncia, Phys. Rev. {\bf D89}, 116011 (2014).
\bibitem{bf11} F. Preis, A. Rebhan and A. Schmitt, J. High Energy Phys. {\bf 1103}, 033 (2011).
\bibitem{db1} E. S. Fraga and A. J. Mizher, Phys. Rev. {\bf D78}, 025016 (2008).
\bibitem{db1sep} E. S. Fraga and A. J. Mizher, Nucl. Phys. {\bf A820}, 103C (2009).
\bibitem{db2} K. Fukushima, M. Ruggieri and R. Gatto, Phys. Rev. {\bf D81}, 114031 (2010).
\bibitem{db3} C. V. Johnson and A. Kundu, J. High Energy Phys. {\bf 12}, 053 (2008).
\bibitem{db5} V. Skokov, Phys. Rev. {\bf D85}, 034026 (2012).
\bibitem{db6} E. S. Fraga, J. Noronha and L. F. Palhares, Phys. Rev. {\bf D87}, 114014 (2013).
\bibitem{pnjl1} R. Gatto and M. Ruggieri, Phys. Rev. {\bf D82}, 054027 (2010).
\bibitem{pnjl1sep} R. Gatto and M. Ruggieri, Phys. Rev. {\bf D83}, 034016 (2011).
\bibitem{pnjl2} M. Ferreira, P. Costa and C. Provid$\hat e$ncia, Phys. Rev. {\bf D89}, 036006 (2014).
\bibitem{pnjl3} M. Ferreira, P. Costa, D. P. Menezes, C. Provid$\hat e$ncia and N.N.Scoccola, Phys. Rev. {\bf D89}, 016002 (2014).
\bibitem{pnjl4} P. Costa,  M. Ferreira,  H. Hansen, D. P. Menezes and C. Provid$\hat e$ncia, Phys. Rev. {\bf D89}, 056013 (2014).
\bibitem{pqm} A. J. Mizher, M. N. Chernodub and E. S. Fraga, Phys. Rev. {\bf D82}, 105016 (2010).
\bibitem{t0effect} E. S. Fraga, B. W. Mintz and J. Schaffner-Bielich, Phys. Lett {\bf B731}, 154-158 (2014).
\bibitem{ferr1} E. J. Ferrer, V.de la Incera, I. Portillo and M. Quiroz, Phys. Rev. {\bf D89}, 085034 (2014).
\bibitem{ferr2} E. J. Ferrer, V.de la Incera, and X. J. Wen, Phys. Rev. {\bf D91}, 054006 (2015).
\bibitem{meimao1} J. Mei and S. J. Mao, Phys. Rev. {\bf D102}, 114035 (2020).
\bibitem{mhuang} K. Xu, J. Y. Chao and M. Huang, Phys. Rev. {\bf D103}, 076015 (2021).
\bibitem{meihuangmao} J. Mei, R. Wen, S. J. Mao, M. Huang and K. Xu, Phys. Rev. {\bf D110}, 034024 (2024).
\bibitem{t0effectmao} S. J. Mao, Phys. Rev. {\bf D110}, 054002 (2024).
\bibitem{maoxn2} S. J. Mao, Chin. Phys. {\bf C49}, 063106 (2025).
\bibitem{maoxn3} S. J. Mao and S. Yang, Phys. Rev. {\bf D112}, 014026 (2025).


\bibitem{c1} N. Agasian and I. Shushpanov, J. High Energy Phys. {\bf 10} (2001) 006.
\bibitem{c3} G. Colucci, E. Fraga and A. Sedrakian, Phys. Lett. \textbf{B}{\bf 728}, 19 (2014).
\bibitem{hadron1} J. Anderson, J. High Energy Phys. 10 (2012) 005, Phys. Rev. \textbf{D}{\bf 86}, 025020 (2012).
\bibitem{hadron2} K. Kamikado and T. Kanazawa, J. High Energy Phys. {\bf 03} (2014) 009.
\bibitem{qm1} G. Krein,and C. Miller, {\it Symmetry}, 13(4): 551 (2021).
\bibitem{qm2}A. Ayala, J. L. Hern$\acute{a}$ndez, L. A. Hern$\acute{a}$ndez, R. L. S. Farias, R. Zamora, Phys. Rev. \textbf{D}\textbf{103}, 054038(2021).
\bibitem{sigma1} R.~M.~Aguirre, Phys.\ Rev.\ \textbf{D}{\bf 96}, 096013 (2017).
\bibitem{sigma2} A.~Ayala, R.~L.~S.~Farias, S.~Hern\'{a}ndez-Ortiz, L.~A.~Hern\'{a}ndez, D.~M.~Paret and R.~Zamora, Phys.\ Rev.\ \textbf{D}{\bf 98}, 114008 (2018).
\bibitem{sigma3} A.~Das and N.~Haque, Phys.\ Rev.\ \textbf{D}{\bf 101}, 074033 (2020).
\bibitem{sigma4} A. N. Tawfik, A. M. Diab and T. M. Hussein, Chin. Phys. {\bf C43}, 034103 (2019).

\bibitem{l1} Y. Hidaka and A. Yamatomo, Phys. Rev. \textbf{D}{\bf 87}, 094502 (2013).
\bibitem{l2} E. Luschevskaya, O. Solovjeva, O. Kochetkov and O. Teryaev, Nucl. Phys. \textbf{B}{\bf 898}, 627 (2015).
\bibitem{l3} E. Luschevskaya, O. Solovjeva and O. Teryaev, Phys. Lett. \textbf{B}{\bf 761}, 393 (2016).
\bibitem{l4} G. S. Bali, B. Brandt, G. Endr\H{o}di and B. Gl\"{a}{\ss}le, Phys. Rev. \textbf{D}{\bf 97}, 034505 (2018).
\bibitem{lqcd5} G. S. Bali, B. Brandt, G. Endr\H{o}di and B. Gl\"{a}{\ss}le, PoS LATTICE {\bf 2015}, 265 (2016).
\bibitem{lqcd6} G. S. Bali, F. Bruckmann, G. Endr\H{o}di, Z. Fodor, S. D. Katz, S. Krieg, A. Sch\"{a}fer and K. K. Szab\'{o}, J. High Energy Phys. {\bf 02} (2012) 044.
\bibitem{ding2008.00493} H. T. Ding, S. T. Li, A. Tomiya, X. D. Wang and Y. Zhang, Phys. Rev. \textbf{D}{\bf 104}, 014505 (2021).

\bibitem{njl2} S. Klevansky, Rev. Mod. Phys. {\bf 64}, 649 (1992).
\bibitem{meson} S. Avancini, R. Farias, M. Pinto, W. Travres and V. Tim$\acute{o}$teo, Phys. Lett. \textbf{B}{\bf 767}, 247 (2017).
\bibitem{mfir} S. Avancini, W. Travres and M. Pinto, Phys. Rev. \textbf{D}{\bf 93}, 014010 (2016).
\bibitem{ritus5} S. Fayazbakhsh, S. Sadeghian and N. Sadooghi, Phys. Rev. \textbf{D}{\bf 86}, 085042 (2012).
\bibitem{ritus6} S. Fayazbakhsh and N. Sadooghi, Phys. Rev. \textbf{D}{\bf 88}, 065030 (2013).
\bibitem{mao1} S. J. Mao, Phys. Lett. \textbf{B} {\bf 758}, 195 (2016).
\bibitem{mao11} S. J. Mao, Phys. Rev. \textbf{D} {\bf 94}, 036007 (2016).
\bibitem{mao2} S. J. Mao and Y. X. Wang, Phys. Rev. \textbf{D}{\bf 96}, 034004 (2017).
\bibitem{wang} Z. Y. Wang and P. F. Zhuang, Phys. Rev. \textbf{D}{\bf 97}, 034026 (2018).
\bibitem{coppola} M. Coppola, D. Dumm and N. Scoccola, Phys. Lett \textbf{B}{\bf 782}, 155-161 (2018).
\bibitem{phi} R. Zhang, W. J. Fu and Y. X. Liu, J. Eur. Phys. \textbf{C}{\bf 76}, 307 (2016).
\bibitem{liuhao3} H. Liu, X. Wang, L. Yu and M. Huang, Phys. Rev. \textbf{D}{\bf 97}, 076008 (2018).
\bibitem{he} D. N. Li, G. Q. Cao and L. Y. He, Phys. Rev. \textbf{D}{\bf 104}, 074026 (2021).
\bibitem{maocharge} S. J. Mao, Phys. Rev. \textbf{D}\textbf{99}, 056005 (2019).
\bibitem{maopion} L. Y. Li and S. J. Mao, Chin. Phys. \textbf{C46}, 094105 (2022).
\bibitem{yulang2010.05716} B. K. Sheng, Y. Y. Wang, X. Y. Wang, L. Yu, Phys. Rev. \textbf{D}{\bf 103}, 094001 (2021).
\bibitem{yulang2} B. K. Sheng, X. Y. Wang, L. Yu, Phys. Rev. \textbf{D}{\bf 105}, 034003 (2022).
\bibitem{q1} D. G. Dumm, M. I. Villafa\~{n}e and N. N. Scoccola, Phys. Rev. \textbf{D}{\bf 97}, 034025 (2018).
\bibitem{q2} S.~S.~Avancini, R.~L.~S.~Farias and W.~R.~Tavares, Phys.\ Rev.\ \textbf{D}{\bf 99}, 056009 (2019).
\bibitem{q3} N.~Chaudhuri, S.~Ghosh, S.~Sarkar and P.~Roy, Phys.\ Rev.\ \textbf{D}{\bf 99}, 116025 (2019).
\bibitem{q4} M.~Coppola, D.~G. Dumm, S.~Noguera and N.~N.~Scoccola, Phys.\ Rev.\ \textbf{D}{\bf 100}, 054014 (2019).
\bibitem{huangamm1} J. Y. Chao, Y. X. Liu and L. Chang, arXiv:2007.14258.
\bibitem{huangamm2} K. Xu, J.Y. Chao and M. Huang, Phys.\ Rev.\ \textbf{D}{\bf 103}, 076015 (2021).
\bibitem{q5}  V.~D.~Orlovsky and Y.~A.~Simonov, J. High Energy Phys. {\bf 09} (2013) 136.
\bibitem{q6} K.~Hattori, T.~Kojo and N.~Su, Nucl.\ Phys.\ \textbf{A} {\bf 951}, 1 (2016).
\bibitem{q7} M.~A.~Andreichikov, B.~O.~Kerbikov, E.~V.~Luschevskaya, Y.~A.~Simonov and O.~E.~Solovjeva, J. High Energy Phys. {\bf 05} (2017) 007.
\bibitem{q8} Y.~A.~Simonov, Phys.\ Atom.\ Nucl.\  {\bf 79}, 455 (2016). [Yad.\ Fiz.\  {\bf 79}, 277 (2016)].
\bibitem{q9} M.~A.~Andreichikov and Y.~A.~Simonov, Eur.\ Phys.\ J. \textbf{C}{\bf 78}, 902 (2018).
\bibitem{q10} C.~A.~Dominguez, M.~Loewe and C.~Villavicencio, Phys.\ Rev.  {\bf D98}, 034015 (2018).

\bibitem{meijie3njl} J. Mei, T. Xia and S. J. Mao, Phys.\ Rev.  {\bf D107}, 074018 (2023); Phys. Rev.  {\bf D110}, 119901(E) (2024).
\bibitem{feng} C. Y. Yang, S. Q. Feng, Phys.\ Rev.  {\bf D112}, 036008 (2025).

\bibitem{ding2026} H. T. Ding and D. Zhang, arXiv:2601.18354.

\bibitem{su3meson1} T. Kojo, Eur. Phys. J. \textbf{A}{\bf 57}, 317 (2021).
\bibitem{su3meson3} A. Mishra and S. Misra, Int. J. Mod. Phys. \textbf{E}{\bf 30}, 2150014 (2021).
\bibitem{su3meson4} S. S. Avancini, M. Coppola, N. N. Scoccola, and J. C. Sodr\'{e}, Phys. Rev. \textbf{D}{\bf 104}, 094040 (2021).


\bibitem{Chernodub:2010qx}
M.~N.~Chernodub,
Phys.\ Rev.\ \textbf{D}{\bf 82}, 085011 (2010).

\bibitem{Chernodub:2011mc}
M.~N.~Chernodub,
Phys.\ Rev.\ Lett.\  {\bf 106}, 142003 (2011).
\bibitem{Callebaut:2011uc}
N.~Callebaut, D.~Dudal and H.~Verschelde,
arXiv:1102.3103.

\bibitem{Ammon:2011je}
M.~Ammon, J.~Erdmenger, P.~Kerner and M.~Strydom,
Phys.\ Lett.\ \textbf{B}{\bf 706}, 94 (2011).

\bibitem{Cai:2013pda}
R.~G.~Cai, S.~He, L.~Li and L.~F.~Li,
J. High Energy Phys. {\bf 12} (2013) 036.




\bibitem{Frasca:2013kka}
M.~Frasca,
J. High Energy Phys. {\bf 11} (2013) 099.


\bibitem{Andreichikov:2013zba}
M.~A.~Andreichikov, B.~O.~Kerbikov, V.~D.~Orlovsky and Y.~A.~Simonov,
Phys.\ Rev.\ \textbf{D}{\bf 87}, 094029 (2013).



\bibitem{Liu:2014uwa}
H.~Liu, L.~Yu and M.~Huang,
Phys.\ Rev.\ \textbf{D}{\bf 91}, 014017 (2015).

\bibitem{Liu:2015pna}
H.~Liu, L.~Yu and M.~Huang,
Chin.\ Phys.\ \textbf{C}{\bf 40}, 023102 (2016).

\bibitem{Liu:2016vuw}
H.~Liu, L.~Yu, M.~Chernodub and M.~Huang,
Phys.\ Rev.\ \textbf{D}{\bf 94}, 113006 (2016).

\bibitem{Kawaguchi:2015gpt}
M.~Kawaguchi and S.~Matsuzaki,
Phys.\ Rev.\ \textbf{D}{\bf 93}, 125027 (2016).





\bibitem{Ghosh:2016evc}
S.~Ghosh, A.~Mukherjee, M.~Mandal, S.~Sarkar and P.~Roy,
Phys.\ Rev.\ \textbf{D}{\bf 94}, 094043 (2016).

\bibitem{Ghosh:2017rjo}
S.~Ghosh, A.~Mukherjee, M.~Mandal, S.~Sarkar and P.~Roy,
Phys.\ Rev.\ \textbf{D}{\bf 96}, 116020 (2017).


\bibitem{Luschevskaya:2014mna}
O.~Larina, E.~Luschevskaya, O.~Kochetkov and O.~V.~Teryaev,
PoS LATTICE {\bf 2014}, 120 (2014).

\bibitem{Luschevskaya:2015bea}
E.~V.~Luschevskaya, O.~A.~Kochetkov, O.~V.~Teryaev and O.~E.~Solovjeva,
JETP Lett.\  {\bf 101}, 674 (2015).



\bibitem{Ding:2020jui}
H.~T.~Ding, S.~T.~Li, S.~Mukherjee, A.~Tomiya and X.~D.~Wang,
PoS LATTICE {\bf 2019}, 250 (2020).



\bibitem{Ghosh}S. Ghosh, A. Mukherjee, N. Chaudhuri, P. Roy and S. Sarkar, Phys. Rev. \textbf{D}\textbf{101}, 056023 (2020).

\bibitem{houphi} X. L. Sheng, S. Y. Yang, Y. L. Zou and D. F. Hou, Eur. Phys. J. {\bf C84}, 299 (2024).
\bibitem{Marasinghe:2011bt}
K.~Marasinghe and K.~Tuchin,
Phys.\ Rev.\ \textbf{C}{\bf 84}, 044908 (2011).

\bibitem{Machado:2013rta}
C.~S.~Machado, F.~S.~Navarra, E.~G.~de Oliveira, J.~Noronha and M.~Strickland,
Phys.\ Rev.\ \textbf{D}{\bf 88}, 034009 (2013).

\bibitem{Alford:2013jva}
J.~Alford and M.~Strickland,
Phys.\ Rev.\ \textbf{D}{\bf 88}, 105017 (2013).

\bibitem{Machado:2013yaa}
C.~S.~Machado, S.~I.~Finazzo, R.~D.~Matheus and J.~Noronha,
Phys.\ Rev.\ \textbf{D}{\bf 89}, 074027 (2014).

\bibitem{Cho:2014exa}
S.~Cho, K.~Hattori, S.~H.~Lee, K.~Morita and S.~Ozaki,
Phys.\ Rev.\ Lett.\  {\bf 113}, 172301 (2014).

\bibitem{Cho:2014loa}
S.~Cho, K.~Hattori, S.~H.~Lee, K.~Morita and S.~Ozaki,
Phys.\ Rev.\ \textbf{D}{\bf 91}, 045025 (2015).

\bibitem{Dudal:2014jfa}
D.~Dudal and T.~G.~Mertens,
Phys.\ Rev.\ \textbf{D}{\bf 91}, 086002 (2015).


\bibitem{Bonati:2015dka}
C.~Bonati, M.~D'Elia and A.~Rucci,
Phys.\ Rev.\ \textbf{D}{\bf 92}, 054014 (2015).


\bibitem{Gubler:2015qok}
P.~Gubler, K.~Hattori, S.~H.~Lee, M.~Oka, S.~Ozaki and K.~Suzuki,
Phys.\ Rev.\ \textbf{D}{\bf 93}, 054026 (2016).


\bibitem{Yoshida:2016xgm}
T.~Yoshida and K.~Suzuki,
Phys.\ Rev.\ \textbf{D}{\bf 94}, 074043 (2016).

\bibitem{Reddy:2017pqp}
Reddy P., Sushruth and Jahan C. S., Amal and Dhale, Nikhil and Mishra, Amruta and J. Schaffner-Bielich,
Phys.\ Rev.\ \textbf{C}{\bf 97}, 065208 (2018).


\bibitem{CS:2018mag}
A.~Mishra, A.~Jahan CS, S.~Kesarwani, H.~Raval, S.~Kumar and J.~Meena,
Eur.\ Phys.\ J.\ \textbf{A}{\bf 55}, 99 (2019).


\bibitem{Tiburzi:2008ma}
B.~C.~Tiburzi,
Nucl.\ Phys.\ \textbf{A}{\bf 814}, 74 (2008).

\bibitem{Andreichikov:2013pga}
M.~A.~Andreichikov, B.~O.~Kerbikov, V.~D.~Orlovsky and Y.~A.~Simonov,
Phys.\ Rev.\ \textbf{D}{\bf 89}, 074033 (2014).

\bibitem{Tiburzi:2014zva}
B.~C.~Tiburzi,
Phys.\ Rev.\ \textbf{D}{\bf 89}, 074019 (2014).


\bibitem{Haber:2014zba}
A.~Haber, F.~Preis and A.~Schmitt,
AIP Conf.\ Proc.\  {\bf 1701}, 080010 (2016).

\bibitem{He:2016oqk}
B.~R.~He,
Phys.\ Lett.\ \textbf{B}{\bf 765}, 109 (2017).

\bibitem{Deshmukh:2017ciw}
A.~Deshmukh and B.~C.~Tiburzi,
Phys.\ Rev.\ \textbf{D}{\bf 97}, 014006 (2018).


\bibitem{Yakhshiev:2019gvb}
U.~Yakhshiev, H.~C.~Kim and M.~Oka,
Phys.\ Rev.\ \textbf{D}{\bf 99}, 054027 (2019).

\bibitem{pnjl5} P. N. Meisinger and M. C. Ogilvie, Phys. Lett. {\bf B379}, 163 (1996).
\bibitem{pnjl6} P. N. Meisinger, T. R. Miller and M. C. Ogilvie, Phys. Rev. {\bf D65}, 034009 (2002).
\bibitem{pnjl7} K. Fukushima, Phys. Lett. {\bf B591}, 277(2004).
\bibitem{pnjl8} A. Mocsy, F. Sannino, and K. Tuominen, Phys. Rev. Lett. {\bf 92}, 182302 (2004).
\bibitem{pnjl9} E. Megias, E. R. Arriola, and L. L. Salcedo, Phys. Rev. {\bf D74}, 065005 (2006).
\bibitem{pnjl10} C. Ratti, M. A. Thaler, and W. Weise, Phys. Rev. {\bf D73}, 014019 (2006).
\bibitem{pnjl12} S. K. Ghosh, T. K. Mukherjee, M. G. Mustafa and R. Ray, Phys. Rev. {\bf D73}, 114007 (2006).

\bibitem{tHooft4} G. 't Hooft, Phys. Rev. \textbf{D}{\bf 14}, 3432 (1976).
\bibitem{tHooft5} G. 't Hooft, Phys. Rept. {\bf 142}, 357 (1986).
\bibitem{tHooft1} T. Kunihiro and T. Hatsuda, Phys. Lett. \textbf{B}{\bf 206}, 385 (1988).
\bibitem{tHooft2} V. Bernard, R. L. Jaffe, and U. G. Meissner, Nucl. Phys. \textbf{B}{\bf 308}, 753 (1988).
\bibitem{tHooft3} H. Reinhardt and R. Alkofer, Phys. Lett. \textbf{B}{\bf 207}, 482 (1988).


\bibitem{njl1} Y. Nambu and G. Jona-Lasinio, Phys. Rev. {\bf 122}, 345 (1961) and {\bf 124}, 246 (1961).
\bibitem{njll2} M. Volkov, Phys. Part. Nucl. {\bf 24}, 35 (1993).
\bibitem{njl3} T. Hatsuda and T. Kunihiro,	Phys. Rep. {\bf 247}, 221 (1994).
\bibitem{njl4} M. Buballa, Phys. Rep. {\bf 407}, 205 (2005).
\bibitem{njl5} P. Rehberg, S. P. Klevansky, and J. H\"{u}fner, Phys. Rev. \textbf{C}{\bf 53}, 410 (1996).


\bibitem{zhuang}P. Zhuang, J. H$\ddot {u}$fner, S.P. Klevansky, Nucl. Phys. {\bf A576}, 525 (1994).

\bibitem{mott1}N. F. Mott, Rev. Mod. Phys. {\bf 40}, 677 (1968).
\bibitem{mott2}J.  H$\ddot {u}$fner,  S.  Klevansky,  and  P.  Rehberg, Nucl.  Phys.  \textbf{A}{\bf 606}, 260 (1996).
\bibitem{mott3}P. Costa, M. Ruivo, and Y. Kalinovsky, Phys. Lett.  \textbf{B}{\bf 560}, 171 (2003).

\end{thebibliography}
\end{document}